\documentclass[12pt,preprint]{aastex}

\shorttitle{Resonant absorption in multi-stranded coronal loops}
\shortauthors{Terradas et al.}

\begin{document}

\title{Resonant absorption in complicated plasma configurations:\\applications
to multi-stranded coronal loop oscillations}

\author{J. Terradas, I. Arregui, R. Oliver, J. L. Ballester}

\affil{Departament de F\'\i sica, Universitat de les Illes Balears, 
E-07122 Palma de Mallorca, Spain}
\and
\author{J. Andries, M. Goossens}

\affil{Centre for Plasma Astrophysics, Katholieke Universiteit Leuven, Celestijnenlaan
200B, B-3001 Leuven, Belgium}

\email{jaume.terradas@uib.es}

\begin{abstract}

We study the excitation and damping of transverse oscillations in a
multi-stranded model of a straight line-tied coronal loop. The transverse geometry of our
equilibrium configuration is quite irregular and more realistic than the usual
cylindrical loop model. By numerically solving the time-dependent ideal
magnetohydrodynamic equations in two dimensions we show how the global motion of
the whole bundle of strands, excited by an external disturbance, is converted
into localized Alfv\'enic motions due to the process of resonant absorption.
This process produces the attenuation of the transverse oscillations. At any location in the structure two dominant frequencies are
found, the frequency of the global mode, or quasi-mode, and the local Alfv\'en
frequency. We find that the mechanism of mode conversion, due to the coupling
between fast and Alfv\'en waves, is not compromised by the complicated geometry
of the model. We also show that it is possible to have energy conversion not
only at the external edge of the composite loop but also inside the structure.
The implications of these results and their relationship with the observations
are discussed.

\end{abstract}


\section{Introduction}

The resonant coupling of fast magnetohydrodynamic (MHD) waves and Alfv\'en waves
has been an active research topic for many years. The most important
applications are to laboratory plasmas, the solar corona, and pulsations in the
Earth's magnetosphere. In the present paper we concentrate on the application of
resonant absorption as a possible candidate to explain the damping of transverse
coronal loop oscillations \citep[see][]{hollyang88, gooss92, rudrob02, terr06a}.
The attenuation of the loop oscillations, interpreted as the kink modes, is due
to the energy conversion from the global large scale motions to localized
Alfv\'en modes.

The time-dependent resonant absorption mechanism has been traditionally analyzed
for both driven and initial value problems using different equilibrium models.
For the driven problem this mechanism has been studied in smooth transition
layers \citep{holl87,davila87,hollyang88}, in slab models
\citep{steindav93,ofetal94,ofdav95,ofetal95,degr00,degr02}, and in cylindrical
tubes \citep{grossmith88,poed89,poed90,poedtsker91,saku91,gooss95}. For the
initial value problem similar studies have also focused on smooth interfaces
\citep{ion78,leerob86}, on slabs \citep{steindav93}, and cylinders
\citep{rudrob02,terr06a}. The effects of background flows have been investigated
by  \citet{hollw90,erd96,tirry98,and00,and01}.

The time-dependent results have been complemented by eigenmode calculations
\citep[see for example][]{tirry96}. The coupling of the fast kink mode with the
Alfv\'en continuum produces the damping of the oscillations. In this case the
global mode is usually called the quasi-mode and has a complex frequency. The
theoretical models that have been used to calculate the quasi-modes are so far
simple but necessary to understand the main properties of resonant absorption.
Recently, new effects have been investigated. The curvature effect has been
studied by \citet{vand04b} and \citet{terr06b}, and the influence of
stratification along the loop has been analyzed by \citet{and05} and by
\citet{arr05}. The effect of non-circularity of the tube cross-section has been
analyzed by \citet{rud03}, while the effect of the internal structure in loops,
using a slab model, has been studied by \citet{arr07}. The overall conclusion of
these recent investigations is that these new ingredients do not significantly
change the damping per period of the oscillations. This suggests that resonant
absorption is a robust mechanism and that its efficiency is not easily affected
by the considered effects.

Nevertheless, even the most recent models are still too simple compared with the
real conditions in coronal loops. For example, there are no physical reasons to
think that the cross-section of coronal loops is perfectly cylindrical.
Moreover,  there is observational information suggesting that loops are not
monolithic (as they are usually modeled) but that they are formed by bundles of
individual strands considered as mini-loops for which the heating and plasma
properties are approximately uniform in the transverse direction
\citep{schmelz}. This view is supported by some authors \citep{martens,klim} but
not by others \citep{asch05,aschnig05}. Thus, one of the questions that arise
from these observations is how the internal structure of coronal loops can
modify the mechanism of resonant absorption.

The purpose of this paper is to study the mechanism of resonant absorption and
its efficiency in a rather arbitrary two-dimensional distribution of plasma, and
to analyze the implications on the damping of transversal coronal loop
oscillations in a multi-stranded loop model. Instead of calculating the
eigenmodes of the structure, which is difficult due to the complicated geometry,
we investigate the dynamical response of the bundle of individual strands to an
initial perturbation. Thus, we study the initial value problem of the excitation
of the composite loop by solving the time-dependent two-dimensional ideal MHD
equations. The equations are solved numerically using an appropriate code. From
the simulations we study the motions in the loop structure and analyze how the
energy from the global mode is converted into localized Alfv\'enic motions in
the inhomogeneous regions of the loop.

This paper is organized as follows. In \S~\ref{model} the multi-stranded loop
model is described and the basic MHD equations are given. The numerical method
used to solve the time-dependent equations is explained in \S~\ref{method}. In
\S~\ref{test} the results of the time-dependent problem for a single cylindrical
inhomogeneous loop are compared with the eigenmode calculations. In
\S~\ref{results} the time-dependent problem for the multi-stranded loop is
considered and several features of the resonant absorption mechanism are
discussed, in particular the behavior in the inhomogeneous layer, and the
energetics of the problem is investigated. Finally, in \S~\ref{concl} the main
conclusions are drawn.

\section{Loop Model and Governing MHD Equations}\label{model}

\begin{figure}[!ht]
\center{

\resizebox{6.5cm}{!}{\includegraphics{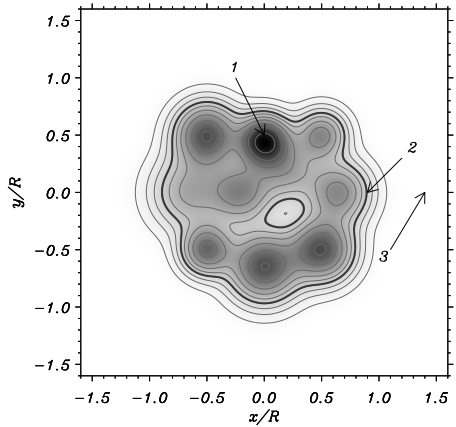}}
}
\caption{ \small 
Cross section of the density of the multi-stranded loop model.  The lengths are
normalized to $R$, the reference radius of an individual homogeneous loop. The
contour lines represent curves of constant Alfv\'en frequency, given by
equation~(\ref{freqAlf}) (they also represent contours of constant density).
The thick line corresponds to the Alfv\'en frequency that matches the frequency
of the global mode. The labels $1$, $2$, and $3$
indicate the locations analyzed in Figures~\ref{signalplotint},
\ref{signalplotres}, and \ref{signalplotex}, respectively.}  \label{density}
\end{figure}

The equilibrium magnetic field is straight, uniform, and pointing in the
$z-$direction, ${\bf B}=B_0\,{\bf {\hat e}_z}$. For applications to the solar
corona, it is a good approximation to consider that the magnetic pressure
dominates the gas pressure. This zero$-\beta$ approximation allows us to choose
an arbitrary  density profile. As a model of a bundle of loops we use a
superposition of $N$ tiny, parallel cylinders with different radii and densities. In
Cartesian coordinates, the cross-section of the density of each individual
strand is assumed to have the following form,

\begin{eqnarray}\label{indvstrand} \rho_i(x,y)=\rho_{0i}\,
\exp\left[{-\frac{(x-x_i)^2+(y-y_i)^2}{a_i^2}}\right],
\end{eqnarray}
where $\rho_{0i}$ is the maximum density of the strand, $x_i$ and $y_i$ the
position of the strand axis, and $a_i$ the strand radius.
The density of the multi-stranded model is defined
as
\begin{eqnarray}\label{totdens} \rho_0(x,y)=\sum^N_{i=1} \rho_{i}(x,y)+\rho_{\rm
ex},
\end{eqnarray}
$\rho_{\rm ex}$ being the density of the external medium (assumed constant).
Note that the density of a single uniform cylindrical loop, $\rho_{\rm in}$, with
radius $R$ that has the same mass as the multi-stranded loop is simply, 
\begin{eqnarray}\label{totdenscyl}
\rho_{\rm in}=\frac{1}{R^2}\sum^N_{i=1}
\rho_{0i}\,a_{i}^2+\rho_{\rm ex}.
\end{eqnarray}
In Figure~\ref{density} the two-dimensional distribution of the density (the
cross section of the composite loop) is plotted for a particular configuration
based on equations~(\ref{indvstrand}) and (\ref{totdens}). For this
particular configuration the loop is composed of ten individual strands with
their axis located at the following $x_i$ and $y_i$ coordinates:  [0, 0, -0.65,
0.65, 0.5, -0.5, -0.5, 0.5, 0.2, -0.2] and [-0.45, 0.65, 0, 0, 0.5, 0.5, -0.5,
-0.5, 0, -0.2]. The radii of the strands ($a_i/R$) are [0.2, 0.275, 0.25,
0.3, 0.2, 0.25, 0.2, 0.25, 0.25, 0.25], and the densities ($\rho_{0i}/\rho_{00}$) are
[0.8, 0.6, 0.4, 0.3, 0.5, 0.6, 0.4, 0.6, 0.4, 0.3]. The external density is
$\rho_{\rm ex}=1/3\,\rho_{00}$, $\rho_{00}$ being a reference density
inside the loop. 

We see that the loop density has an inhomogeneous distribution with quite an
irregular cross section and irregular boundary. This model has a complex
geometry compared with the usual cylindrical or the elliptical tubes usually
used to study loop oscillations. In this so-called ``spaghetti model"
\citep[see][]{bogdan,kepp} the distance between the strands is quite small and a
strong dynamical interaction between them is expected. This kind of model has
been previously considered in the context of scattering and absorption of sound
waves in composite sunspot models.

To study small amplitude perturbations in this equilibrium we use the linearized
ideal MHD equations. The equilibrium depends on $x$ and $y$ but it is
independent of the longitudinal $z-$coordinate. For this reason, we Fourier
analyze in this direction assuming a dependence of the form $e^{-i k_z z}$,
$k_z$ being  the vertical wavenumber. We concentrate on the fundamental mode,
with $k_z=\pi/L$, $L$ being the length of the loop (here we use $L=20R$). Under
these assumptions the MHD equations are,
\begin{eqnarray}\label{mhd_equations_1}
\rho_0 \frac{\partial v_x}{\partial t}+\frac{B_0}{\mu} \left( i k_z
b_x+\frac{\partial b_z}{\partial x}
\right)&=&0~,\\ \label{mhd_equations_2}
\rho_0 \frac{\partial v_y}{\partial t}+\frac{B_0}{\mu} \left( i k_z
b_y+\frac{\partial b_z}{\partial y} \right)&=&0~,\\
\label{mhd_equations_3}
\frac{\partial b_x}{\partial t}+ B_0 i k_z v_x&=&0~,\\\label{mhd_equations_4}
\frac{\partial b_y}{\partial t}+ B_0 i k_z v_y&=&0~,\\\label{mhd_equations_5}
\frac{\partial b_z}{\partial t}+ B_0 \left( \frac{\partial v_x}{\partial
x} + \frac{\partial v_y}{\partial y}\right)&=&0~,
\end{eqnarray}
where $\mathbf{v}= \left(v_x, v_y, 0 \right)$ is the velocity and $\mathbf{b}=
\left(b_x, b_y, b_z \right)$ is the perturbed magnetic field. These equations
can be rewritten as a system of five equations with five real variables using
the transformations $b_x\rightarrow i\,b_x$ and $b_y\rightarrow i\, b_y$
(these variables are shifted $\pi/2$ in the $z$-direction with respect to $v_x$,
$v_y$, and $b_z$). The equations are complemented with an initial perturbation
located in the external medium. For simplicity we assume that it has the following
form
\begin{eqnarray}\label{pert}
v_y(x,y)=v_{y0}\,\exp\left[-\frac{(y-y_0)^2}{w^2}\right],
\end{eqnarray}
where $v_{y0}$ is the amplitude of the perturbation (this parameter is arbitrary
since we are in the linear regime), $y_0$ is the position of the center of the
disturbance, and $w$ is its width (here we use $y_0=3R$ and $w=R$, which means
that the perturbation is located at a distance $3R$ from the center of the loop). All other
MHD variables are initially set to zero. This perturbation is a planar pulse
which produces the excitation of  fast MHD waves that propagate and interact
with the loop structure.

In the excitation of fast MHD waves and specially Alfv\'en waves an important
quantity in our model is the local Alfv\'en frequency, 
\begin{eqnarray}\label{freqAlf}
\omega_{\rm A}(x,y)=k_z v_{\rm A}(x,y)=k_z \frac{B_0}{\sqrt{\mu \rho_0(x,y)}},
\end{eqnarray}
$v_{\rm A}$ being the Alfv\'en speed. 
The Alfv\'en frequency distribution in our equilibrium depends on $x$ and $y$ in a quite
complicated manner, as we can see in Figure~\ref{density} (see the contours of
constant $\omega_{\rm A}$ represented with
continuous lines). The fact that $\omega_{\rm A}$ varies  with position plays a
key role for the resonant conversion of wave energy from global large scale
motions to localized Alfv\'en modes.

\section{Method}\label{method}

From the numerical point of view the problem that we are studying has basically
two difficulties. The first one is related with the small spatial scales that
are generated in the inhomogeneous layers, so we must make sure that we are
properly resolving  these layers. This point is crucial to have the correct
energy conversion rates and in consequence the right damping times of
transversal oscillations. Due to the phase mixing process
\citep[see][]{heyvpri83}, specially important at the inhomogeneous layers, the
maximum time for which the simulations can be run is determined by the phase
mixing length, defined as the length over which the phase of neighboring
Alfv\'en waves differs by $2\pi$  \citep[][]{mann,wright},
\begin{eqnarray}\label{lph} L_{\rm ph}=2\pi/\left(t\, \frac{d\omega_{\rm
A}}{dx}\right). \end{eqnarray} Therefore, the typical spatial lengths decrease
quickly with time. The shortest wavelength that can be resolved numerically with
a uniform grid spacing of $\Delta x$ is $\lambda=2\Delta x$. We can run the
simulations until the phase mixing length is equal to the shortest wavelength
($L_{\rm ph}=\lambda$). Thus, the maximum simulation time is simply
\begin{eqnarray}\label{tmax} t_{\rm max}=\pi/\left(\Delta x\, \frac{d\omega_{\rm
A}}{dx}\right). \end{eqnarray}  We have checked  that this condition is not
violated in the simulations (note that it also has to be  satisfied  in the
$y-$coordinate).

The second difficulty is the effect of the boundaries on the loop dynamics. Even
applying transparent boundary conditions (zero order extrapolation) at the
limits of the computational domain, it is important to locate the boundaries far
enough from the loop. Note that a trapped mode in the loop has always an
evanescent part in the external medium which might be affected by the conditions
that we apply at the boundaries.

To solve these difficulties we have used a resolution sufficiently high to
resolve the different scales (a grid of $4000 \times 4000$ points is sufficient)
and to avoid significant reflections we have located the domain limits far
enough from the loop,  at $x_B=\pm 8R$, $y_B=\pm 8R$. Nevertheless, in order to
better interpret and visualize the results the plots are displayed in a smaller
spatial domain ($[-1.6R,1.6R]\times[-1.6R,1.6R]$).

To numerically solve the time-dependent MHD equations,
equations~(\ref{mhd_equations_1})$-$(\ref{mhd_equations_5}), together with the
initial condition, equation~(\ref{pert}), we use the code CLAWPACK
\citep{leveq}. This code implements a wide class of high-resolution finite
volume methods for solving linear or nonlinear hyperbolic problems. Due to
resolution requirements we have run the parallelized version of the code in a
cluster of computers.

\section{Test case: resonant absorption in an inhomogeneous cylindrical
loop}\label{test}

\begin{figure}[!ht]
\center{
\resizebox{6.5cm}{!}{\includegraphics{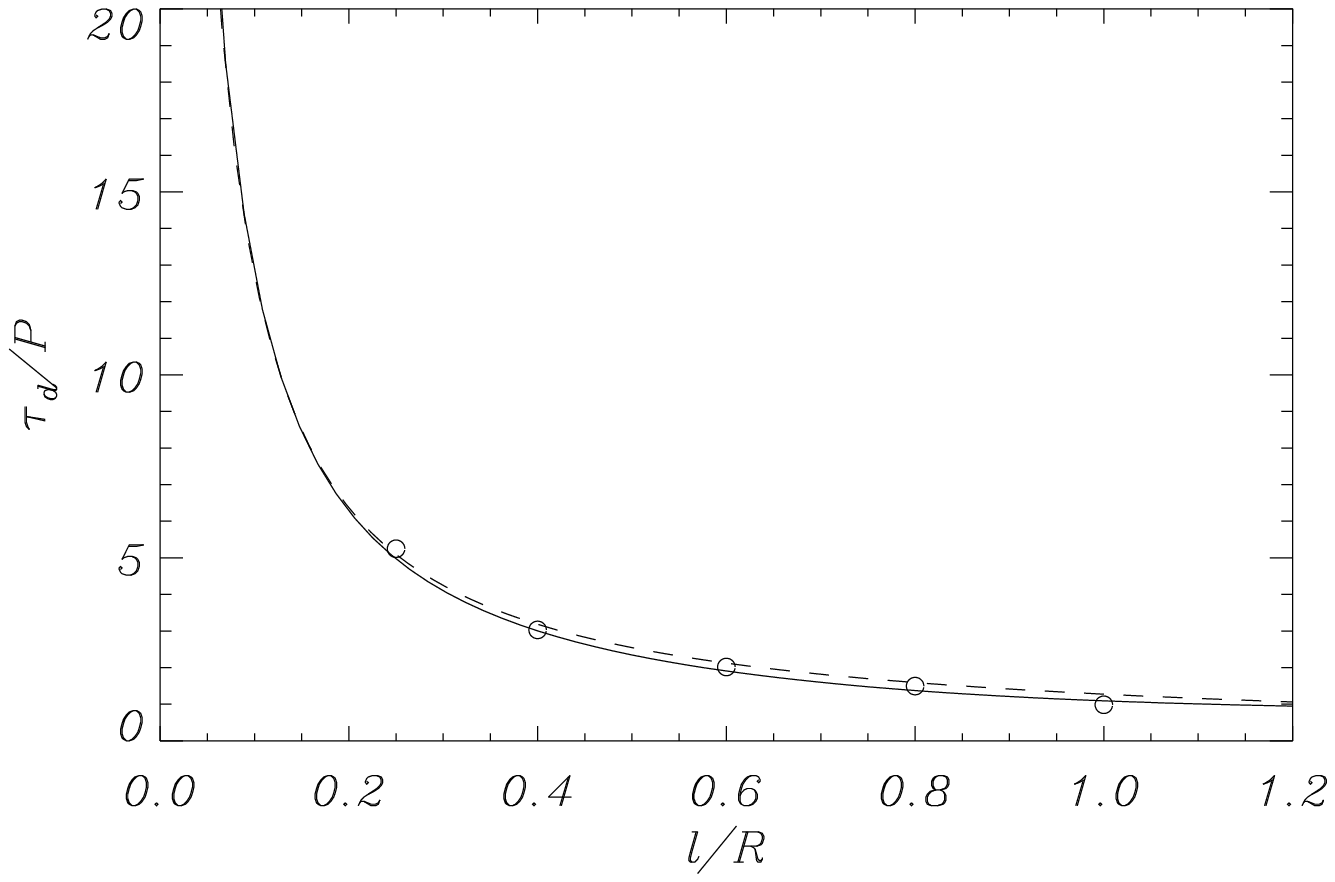}}
\resizebox{6.5cm}{!}{\includegraphics{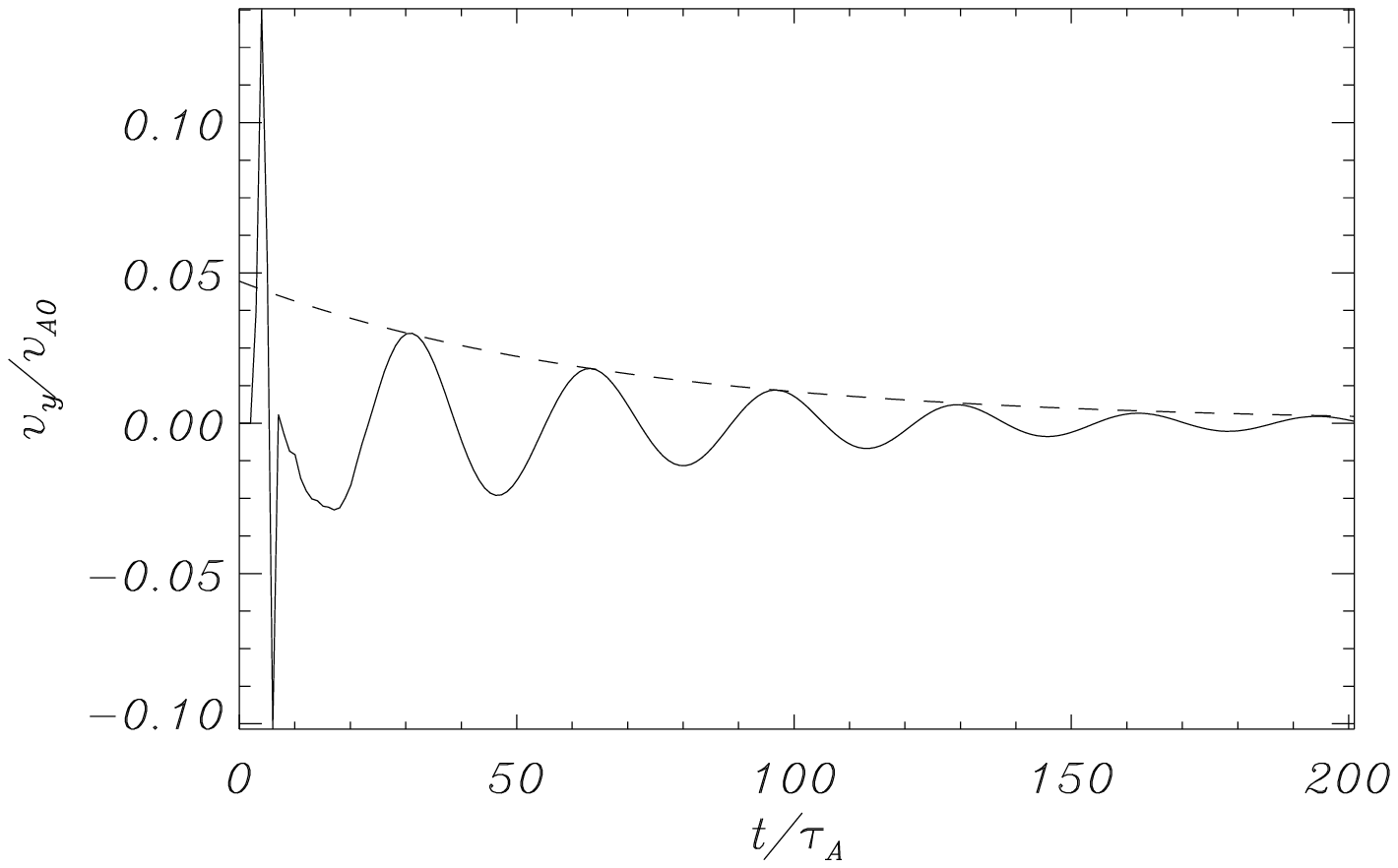}}
}
\caption{\small {\it Top panel:} Damping per period as a function of the thickness of the
non-uniform layer for a single cylindrical loop ($\rho_{\rm ex}=1/3\,\rho_{\rm
in}$, $k_z\,R=\pi/20$). The dashed line is the
damping per period calculated using equation~(\ref{tttb}) while the continuous line is
calculated with the eigenvalue problem for thick layers \citep[see][]{terr06a}. The circles represent
the values calculated from the time-dependent two-dimensional simulations. {\it Bottom panel:} Plot of $v_y$ at the center of the cylindrical
loop as a function of time for the case $l/R=0.6$. This result has been
obtained by solving the time-dependent equations. The dashed line is a fit of
the form $\exp(-t/\tau_d)$. The time is normalized to the Alfv\'en transit time,
$\tau_A=v_{\rm A0}/R$, where  $v_{\rm A0}=B_0/\sqrt{\mu\rho_{\rm in}}$.
} 
\label{tdp} \end{figure}

Due to the different scales involved in the problem the modeling of resonant
absorption is challenging from the numerical point of view. For this reason,
before the ``spaghetti model'' is studied, it is necessary to check that the
numerical method we are using is appropriate to study this problem. As a test we
consider a single inhomogeneous cylindrical loop in two dimensions. To
facilitate comparison of the time-dependent results with previous eigenmode studies we
choose the same density profile as in \citet{rudrob02, vand04a,terr06a}, i.e., a
sinusoidal variation in density across the non-uniform layer. For such a
configuration and for the $m=1$ mode (the kink mode), the damping per period
(the damping time, $\tau_{\rm d}$, over the period, $P$) in the limit of thin tube and
thin boundary is
\begin{eqnarray}\label{tttb}
\frac{\tau_{\rm d}}{P}=\frac{2}{\pi}\frac{R}{l}\frac{\rho_{\rm in}+\rho_{\rm ex}}{\rho_{\rm
in}-\rho_{\rm e}},
\end{eqnarray}
where $l$ is the thickness of the non-uniform layer and $R$ is the loop radius.
For thick layers the eigenvalue problem has to be solved numerically
\citep[see][]{vand04a,terr06a}. In our composed loop model there is a wide range
of thicknesses, and for this reason we have calculated the eigenmodes of the
cylindrical loop numerically \citep[see][for details about the method used to
perform these calculations]{terr06a}. In Figure~\ref{tdp} (top panel) the
damping per period for a single cylindrical loop calculated from
equation~(\ref{tttb}) and the eigenmode calculations are represented as a
function of the thickness of the layer. The small differences between the two
curves are simply due to the fact that equation~(\ref{tttb}) is inaccurate for
thick layers \citep[see also][]{vand04a}. 

To compare the results of these eigenmode calculations with the time-dependent
problem we need to place an initial perturbation in the system that  basically
excites the kink mode (the eigenmode calculations that we want to compare with
are for this mode, $m=1$). The planar pulse perturbation given by
equation~(\ref{pert}) is the kind of disturbance with such a property because an
external perturbation hardly excites high order harmonics such as the fluting
modes, and the sausage modes are leaky in our configuration
\citep[see][]{terr07b}. We have run the code with this perturbation and have
studied the subsequent evolution. Figure~\ref{tdp} (bottom panel) shows the plot
of $v_y$ at the center of the loop as a function of time for a particular
thickness of the layer \citep[see Fig.~4 in][for the analogous simulation in
1D]{terr06a}. After a short transient the loop oscillates with the quasi-mode
period and the amplitude of the mode is attenuated due to the energy conversion
in the inhomogeneous layer. From the results of the simulations we calculate the
period, $P$,  by performing a periodogram. The damping time, $\tau_d$, is
numerically estimated by fitting an exponential function of the form
$\exp(-t/\tau_d)$ to the envelope of the curve. We have performed several
simulations varying the width of the layer and the numerically obtained damping
per periods are represented with circles in Figure~\ref{tdp} (top panel) as a
function of the thickness of the layer. We find the expected dependence with
$l/R$ and we see that the deviations from the eigenmode calculations are quite
small. It is important to remark that, although we model a cylindrical tube, in
these simulations we use Cartesian coordinates and we still obtain the right
damping rates.

In conclusion, with this simple numerical experiment we demonstrate that we are
using a reliable tool to study resonant absorption. Note that contrary to some
precedent works, where a large amount of resistivity is used mainly to avoid
numerical problems, here we are in ideal MHD conditions. Hence, the system
dynamics is not dominated by the resistive regime, in this case the behavior in
Figure~\ref{tdp} (top panel) would be completely different, the damping time
being independent of $l/R$ \citep[see for example Fig.~2 in][]{terr06a}.

\section{Results: multi-stranded model}\label{results}

We now use  the multi-stranded model shown in Figure~\ref{density} and study how
the system evolves due to a planar pulse perturbation (eq.~[\ref{pert}]). The
initial pulse produces a displacement of the whole ensemble of strands basically
in the $y-$direction, i.e. the direction in which the initial perturbation
propagates.  The initial stage of the evolution of the bundle of loops is
dominated by a complicated set of internal reflections of the wavefront between 
the different strands. During this transient several wavefronts propagating from
the bundles into the external medium are found. They correspond to the emission
of the leaky modes representing fast radiating MHD waves with short periods and
fast attenuation \citep[see][for the analysis of such modes in a single
cylindrical loop]{cally86,cally03,terr07a}. After the transient the system
periodically oscillates with certain frequencies. Hereafter, we concentrate on
this stage of the dynamics.

\subsection{Frequency analysis}

\begin{figure}[!ht]
\center{
\resizebox{6.5cm}{!}{\includegraphics{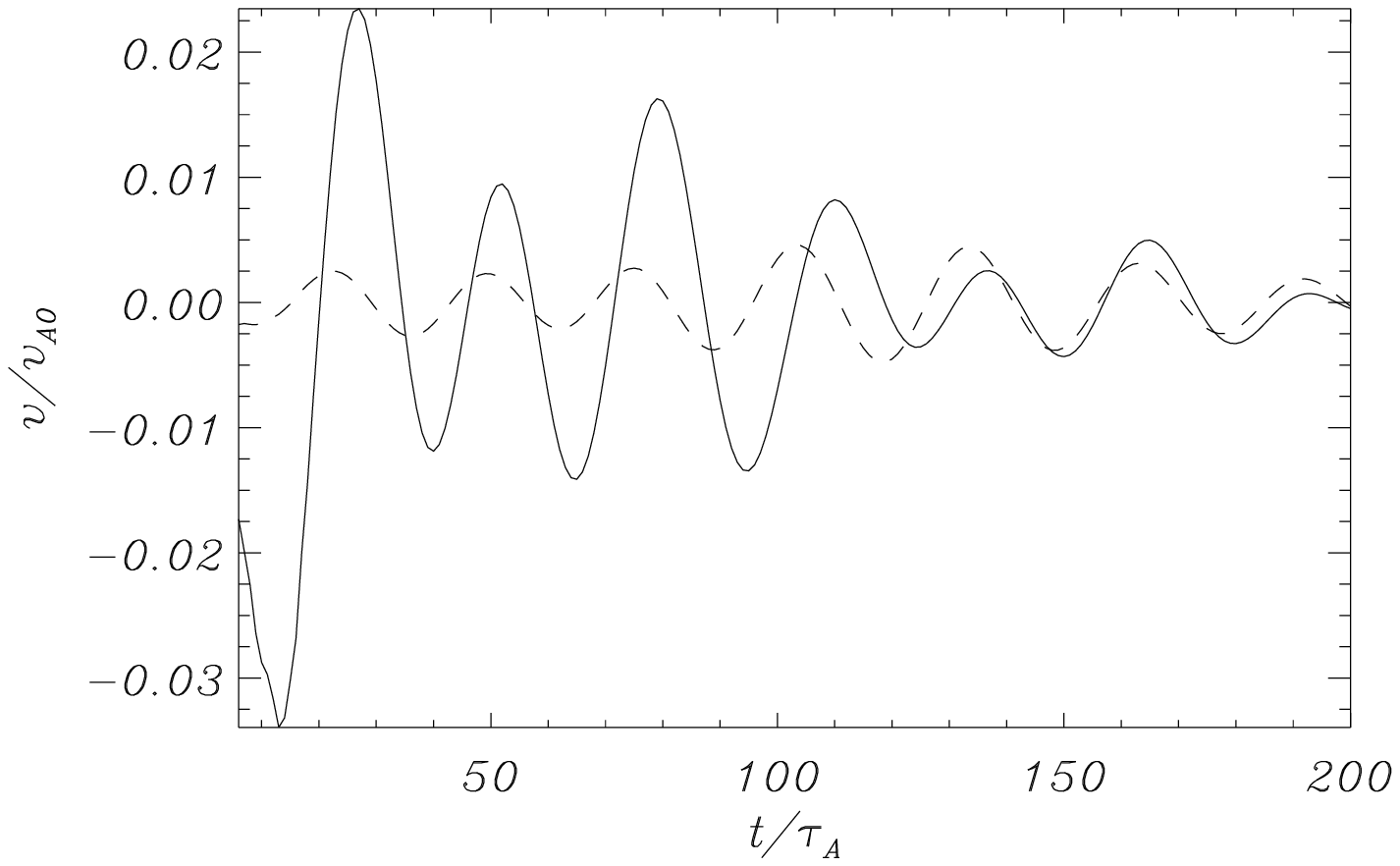}}
\resizebox{6.5cm}{!}{\includegraphics{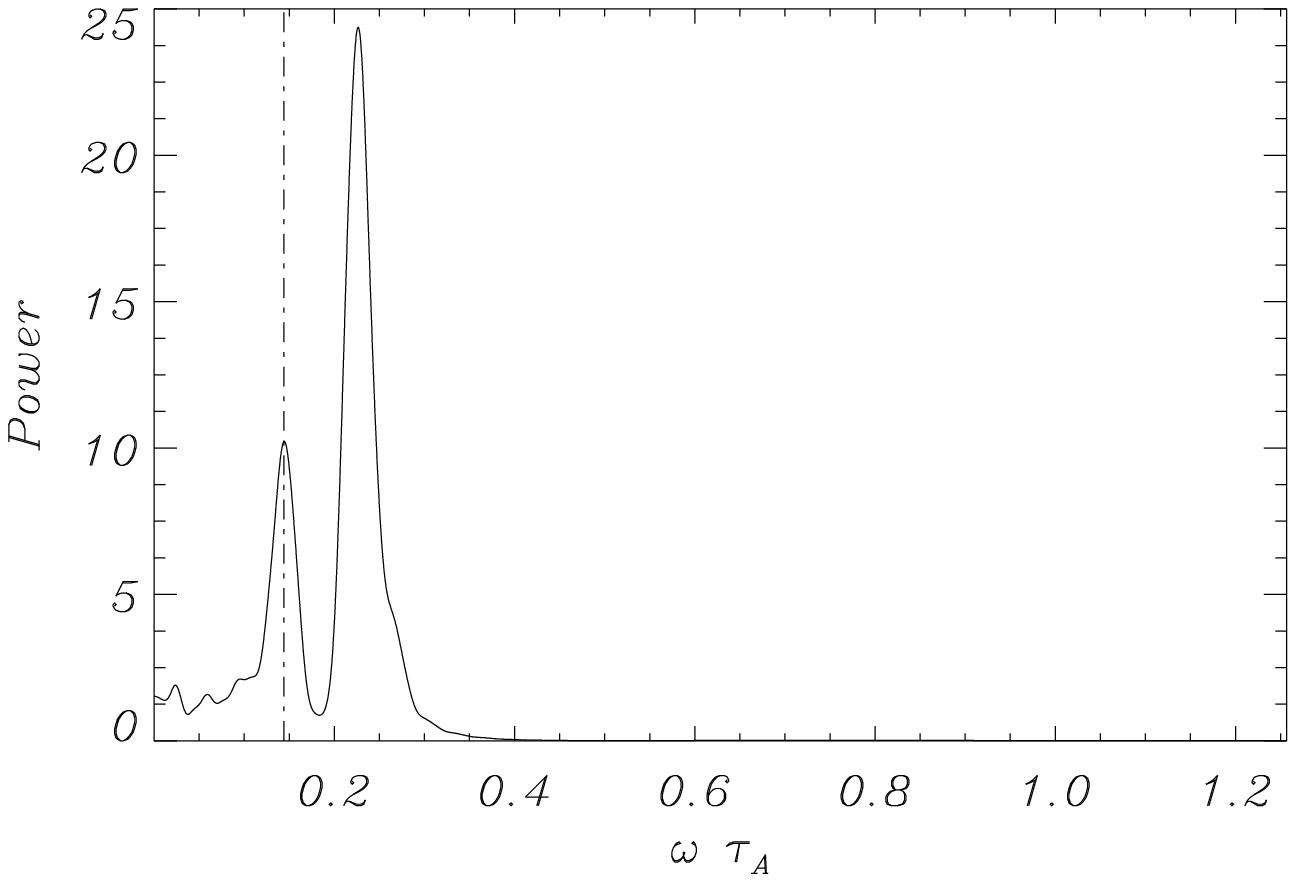}}
}

\caption{ \small  {\it Top panel:} Time evolution of the $v_x$ (dashed line) and
$v_y$ (continuous line) velocity components at the interior point $x=0$, $y=0.5R$
(see label $1$ in Fig.\ref{density}). {\it Bottom panel:} Power spectrum of
the $v_y$ velocity component. The peak with the largest power corresponds to the global mode while the
weaker peak is due to the excitation of the local Alfv\'en mode. The local
Alfv\'en frequency, calculated from the function $\omega_{\rm A}(x,y)$, at the 
coordinates of the previous point is represented with dot-dashed lines and
agrees with the location of the weak peak.}
\label{signalplotint} \end{figure}

We first investigate the characteristic frequencies of the system. An analysis
of the results of the simulations shows that inside the loop there are basically
two dominant frequencies at each point. One is the collective frequency of the
bundle of loops (different from the kink frequency of the individual strands)
and the other is the local Alfv\'en frequency, $\omega_{\rm A}$. The collective
frequency is the result of the excitation of the global mode of the system which
represents the emergent behavior of the entire loop. On the other hand, the
local Alfv\'en frequency is due to the excitation of the Alfv\'en continuum
modes. As an example, in Figure~\ref{signalplotint}  the $v_y$ velocity
component as a function of time (top panel) and its power spectrum (bottom
panel) are represented at an interior point (see label $1$ in
Fig.~\ref{density}). The two dominant frequencies, the local Alfv\'en frequency
and the global frequency (with the largest power) are clearly identified in the
power spectrum. Note the strong decrease of the amplitude with time. This
attenuation  is found almost everywhere at the interior points and suggests a
collective motion of the bundle of tubes. We have also represented in
Figure~\ref{signalplotint} (top panel) the other component of the velocity,
$v_x$. Its amplitude is smaller that the $v_y$ component and it does not show
attenuation with time.

\begin{figure}[!ht]
\center{
\resizebox{6.5cm}{!}{\includegraphics{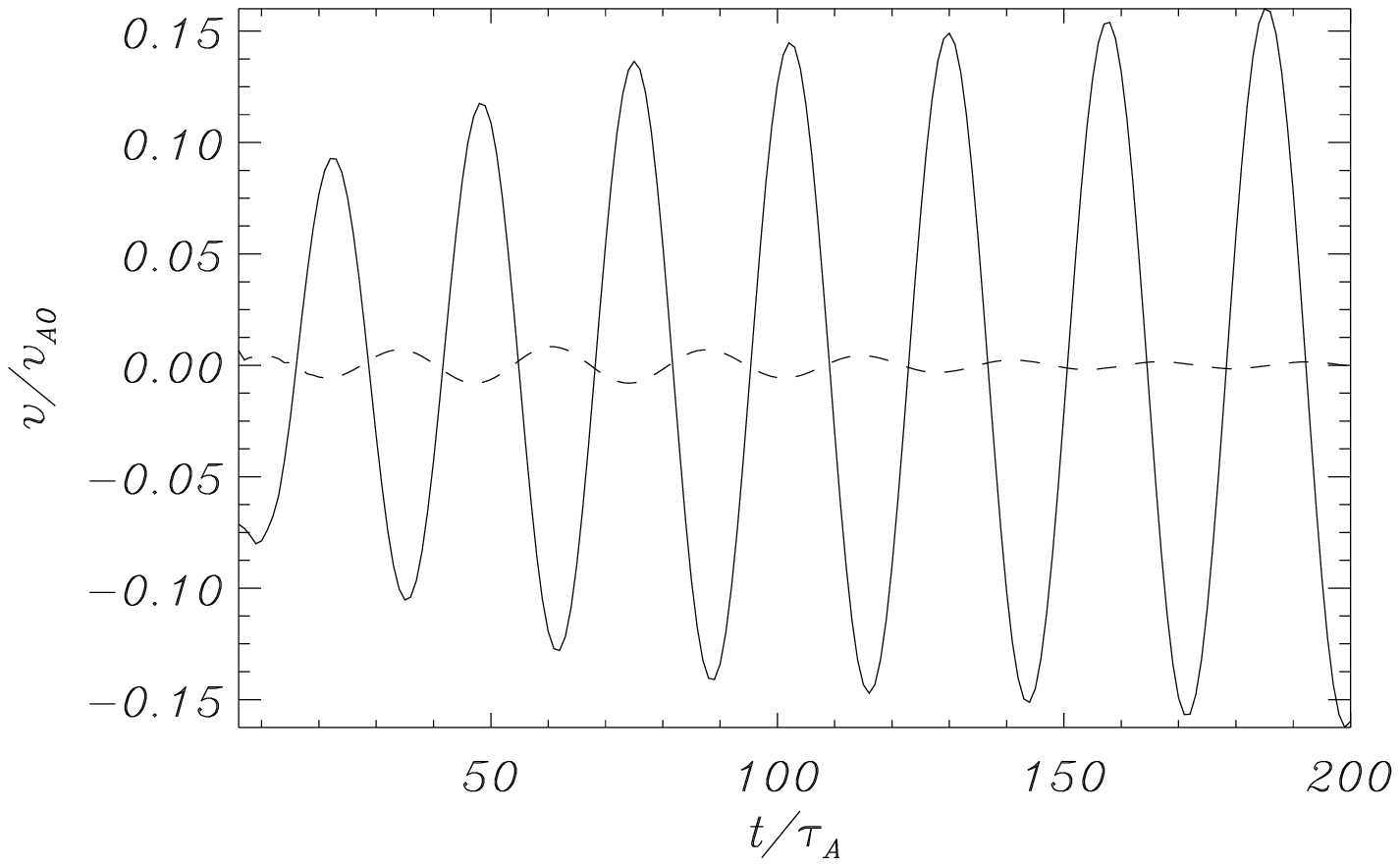}}
\resizebox{6.5cm}{!}{\includegraphics{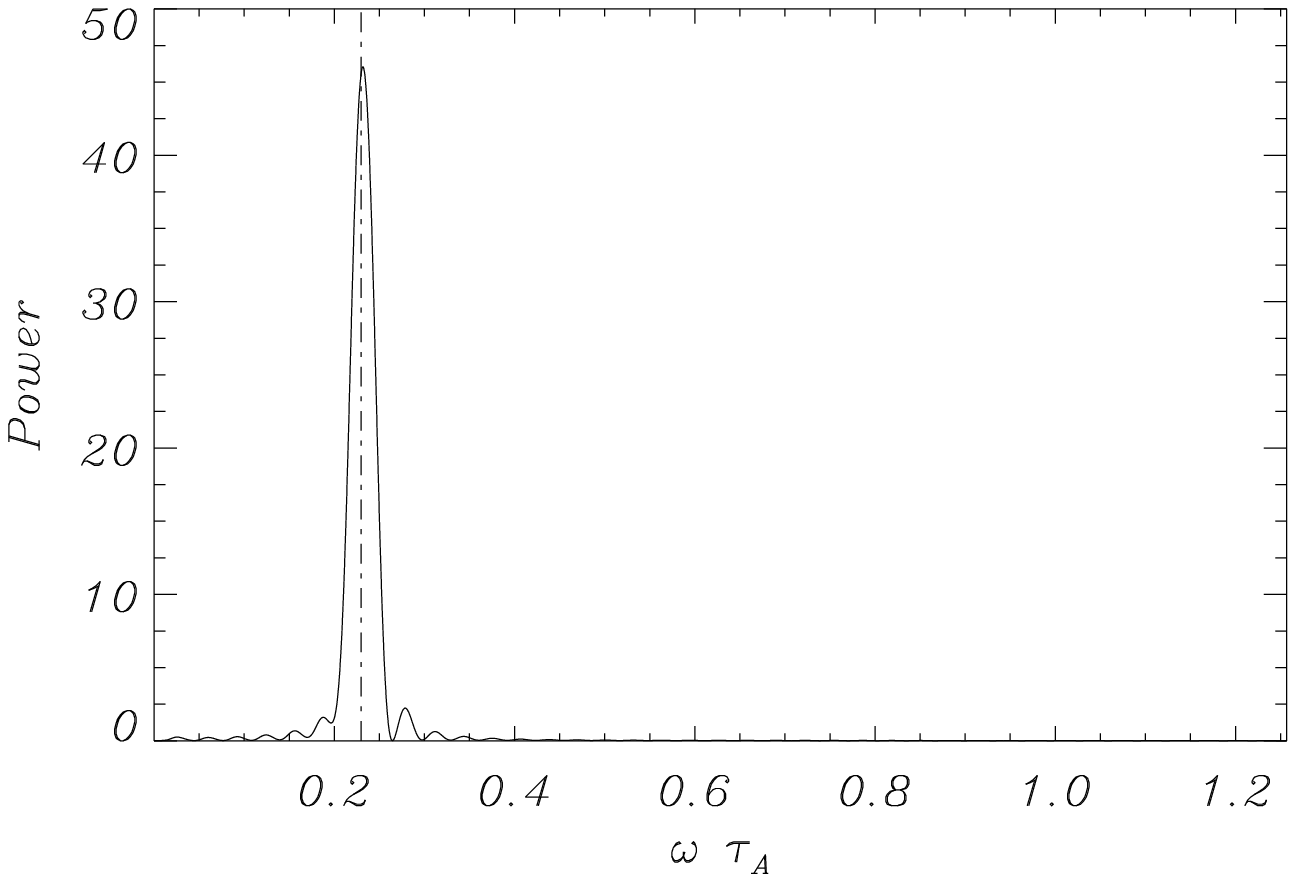}}
}

\caption{ \small  {\it Top panel:} Time evolution of the  $v_x$ (dashed line)
and $v_y$ (continuous line)
component at a position near the loop edge, $x=0.9R$, $y=0$ (see label $2$ in Fig.\ref{density}).
{\it Bottom panel:} Power spectrum of the $v_y$ velocity component. The
local Alfv\'en frequency is represented with
dot-dashed lines.} \label{signalplotres}
\end{figure}

\begin{figure}[!ht]
\center{
\resizebox{6.5cm}{!}{\includegraphics{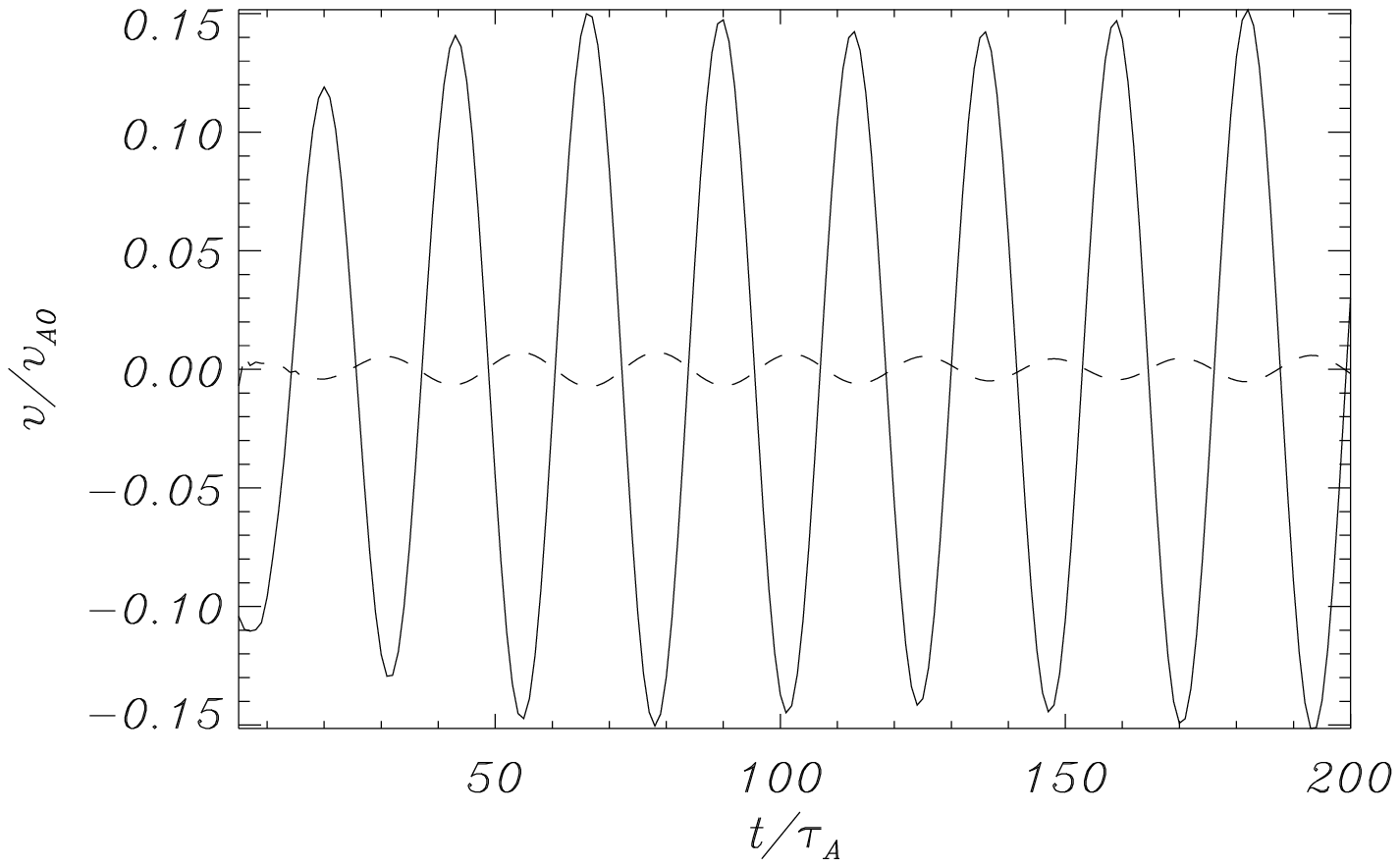}}
\resizebox{6.5cm}{!}{\includegraphics{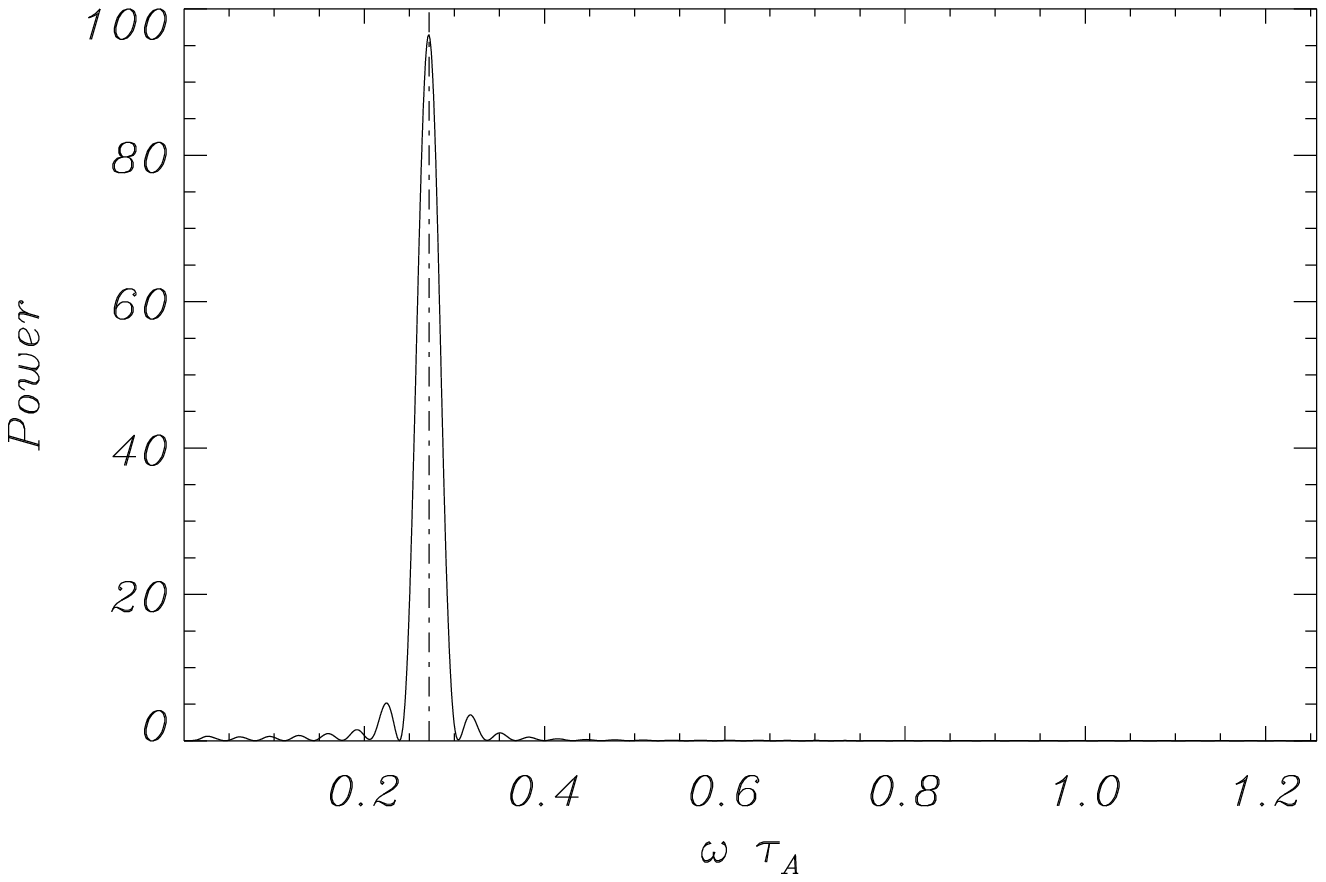}}
}

\caption{ \small  {\it Top panel:} Time evolution of the  $v_x$ (dashed line)
and $v_y$ (continuous line)
component at a position near the loop edge, $x=1.4R$, $y=0$ (see label $3$ in Fig.\ref{density}).
{\it Bottom panel:} Power spectrum of the $v_y$ velocity component. The
local Alfv\'en frequency, i.e. for this
particular point the external Alfv\'en frequency, is represented with
dot-dashed lines.} \label{signalplotex}
\end{figure}

There are particular locations in the structure where the behavior of the signal
is completely different. An example is shown in Figure~\ref{signalplotres} which
corresponds to a position situated near the loop edge (see label $2$ in
Fig.~\ref{density}). Now, instead of a damped oscillation, the amplitude of the
$v_y$ velocity component grows with time while the $v_x$ component decreases.
There is just a single peak in the power spectrum because the local and the
global frequencies coincide. As we will show in the following sections this is
the place where the energy conversion takes place.

Outside the structure, oscillations tending to the Alfv\'en frequency are found
everywhere, see for example the velocity  represented in
Figure~\ref{signalplotex} (see label $3$ in Fig.~\ref{density}). These
oscillations are basically produced by the wake of the initial fast MHD
perturbation and have a large amplitude compared with the amplitude inside the
loop (see Fig.~\ref{signalplotint}). This wake oscillates (in the limit of large
times) at the local Alfv\'en frequency \citep[see][]{terr05} which is precisely
where the dominant peak is found in the power spectrum. However, it is worth
noticing that near the loop edge, but in the external medium, the quasi-mode
frequency is also present. This is not surprising, since the quasi-mode has an
evanescent tail in the external medium.

We have also performed the same analysis but for the $b_z$ component, which is
proportional to the total pressure perturbation ($P_{\rm T}=b_z\,B_0/\mu$). We
have found large power at the global mode frequency and very weak power around
the local Alfv\'en frequency. This is because pure Alfv\'en modes do not perturb
the total pressure and,  although in our configuration  Alfv\'en modes are
coupled with fast modes, they still keep a strong incompressible character.

\subsection{Velocity field}

\begin{figure}[!ht]
\center{
\resizebox{6.cm}{!}{\includegraphics{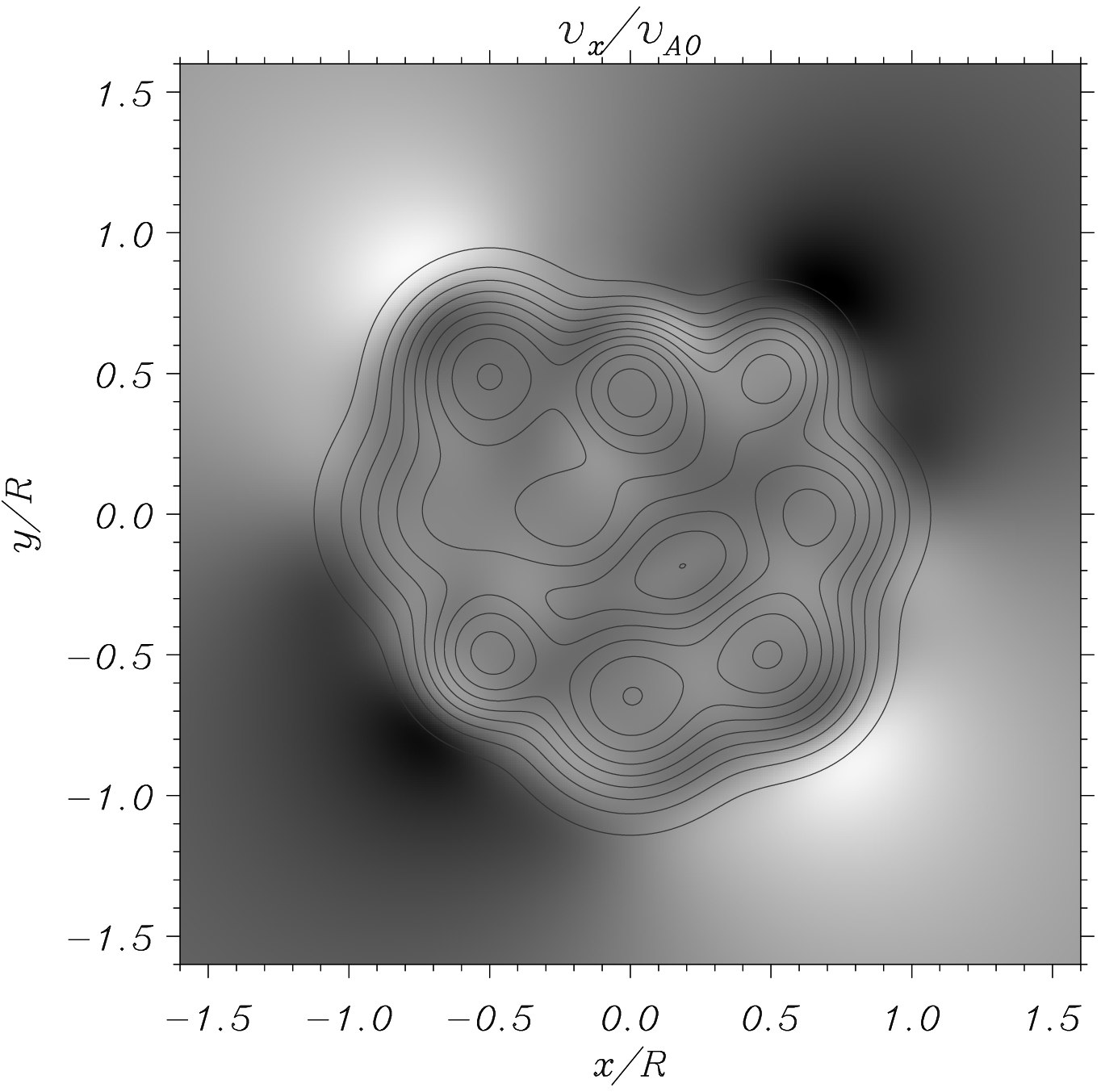}}
\resizebox{6.cm}{!}{\includegraphics{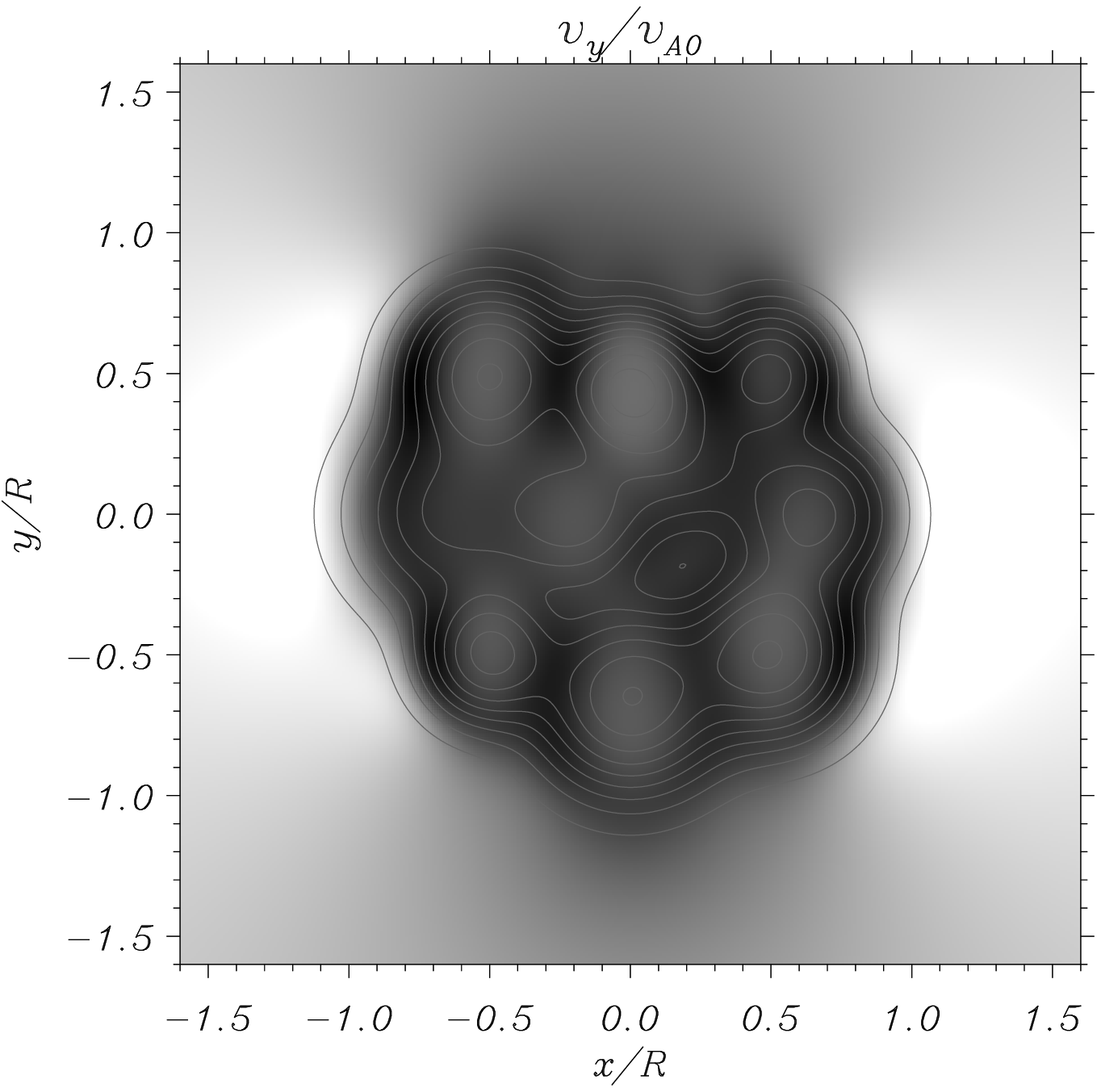}}
}\center{\hspace{0.5cm}
\resizebox{6.cm}{!}{\includegraphics{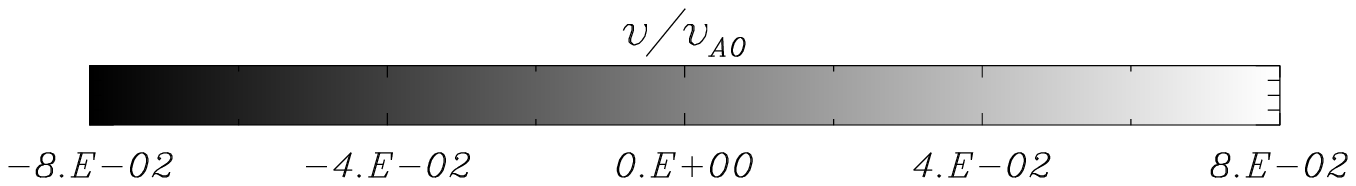}}

} \caption{ \small  Spatial distribution of the $v_x$ velocity component ({\it
top panel}), $v_y$ velocity component ({\it bottom panel}) at $t=40\,\tau_A$.
The corresponding velocity field is plotted in the top panel of
Figure~\ref{resonance}. Contours of constant Alfv\'en frequency are represented
with continuous lines.}  \label{contfields} \end{figure}

\begin{figure}[!ht]
\center{
\resizebox{6.cm}{!}{\includegraphics{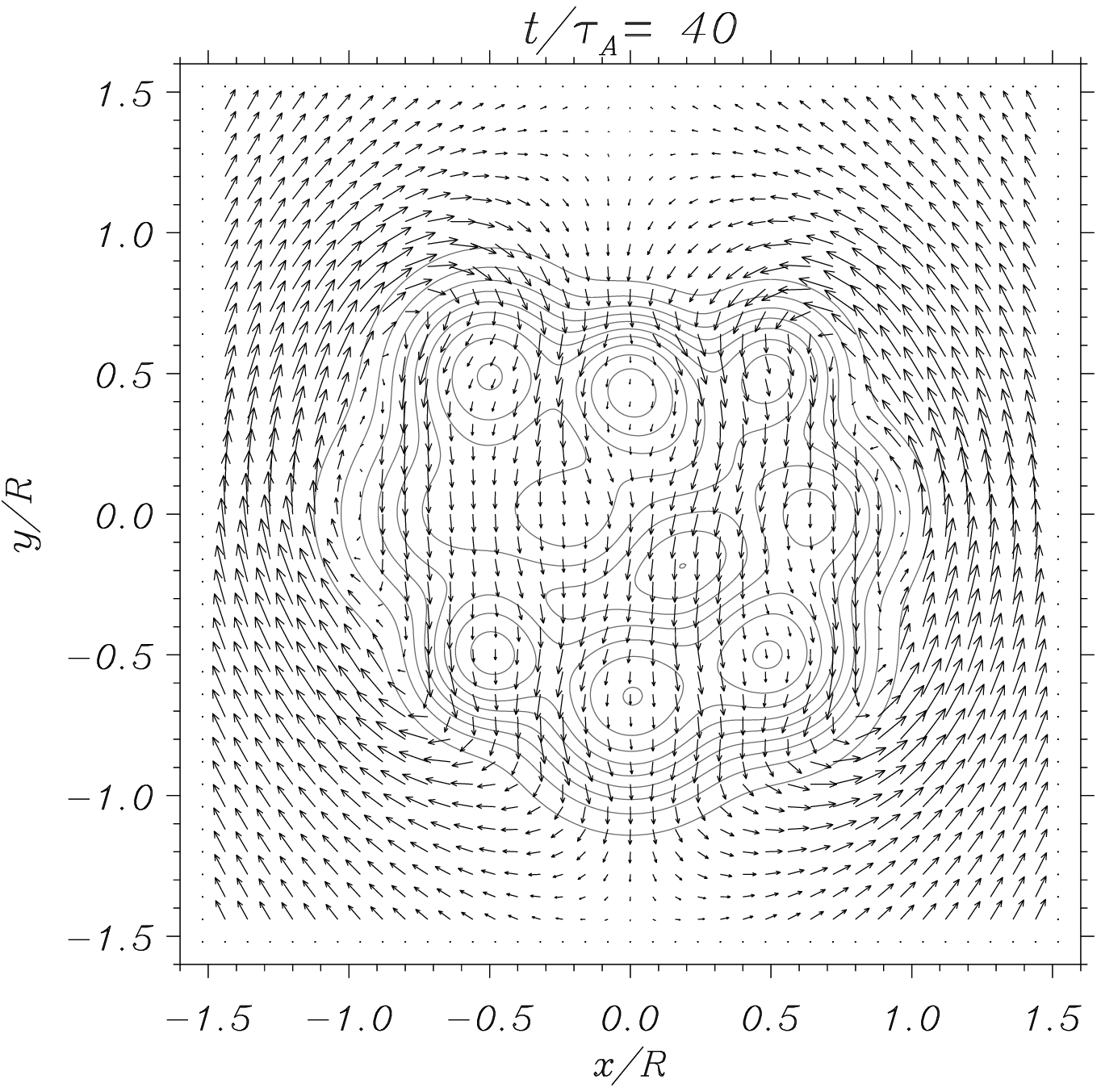}}
\resizebox{6.cm}{!}{\includegraphics{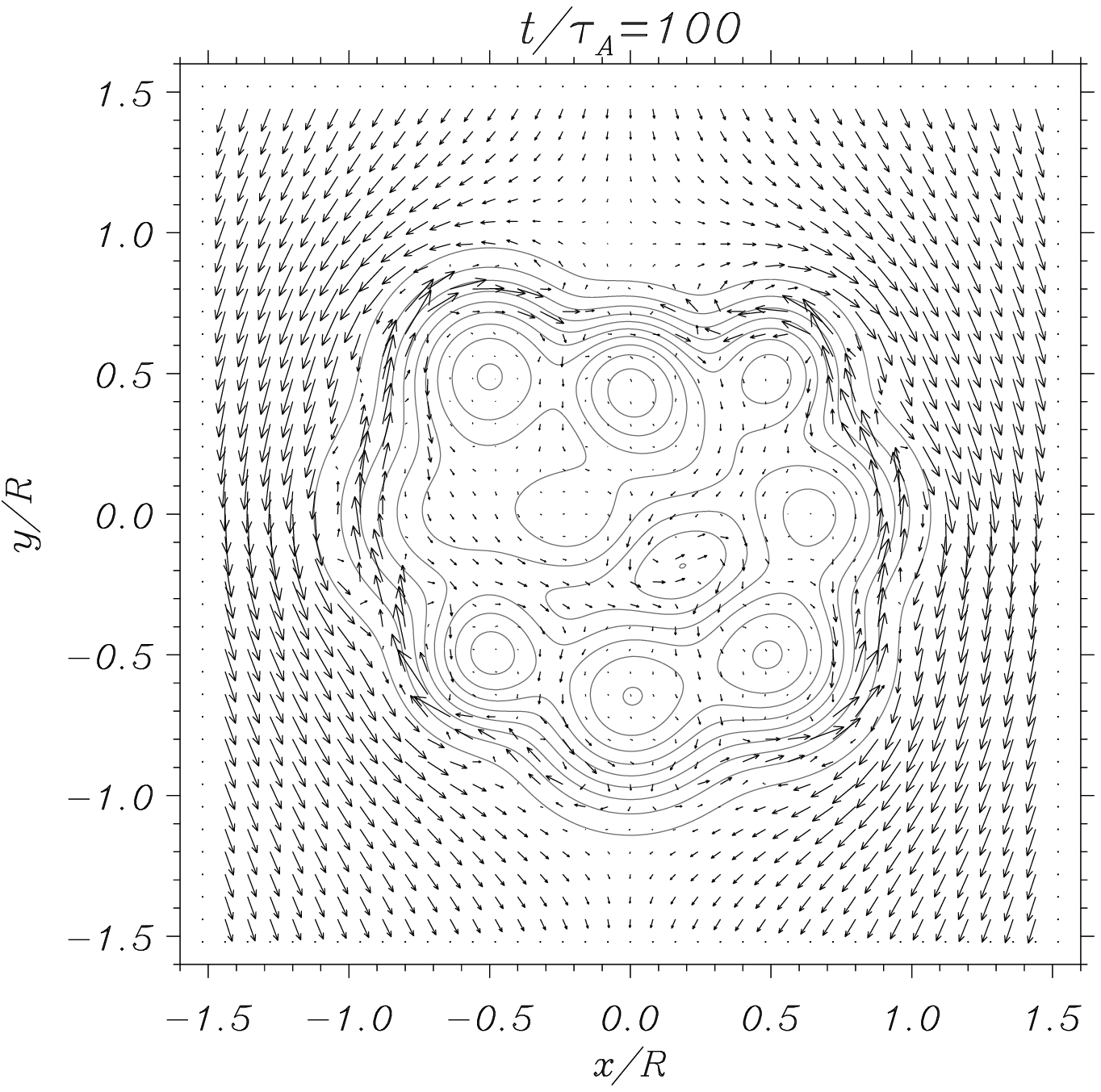}}
\resizebox{6.cm}{!}{\includegraphics{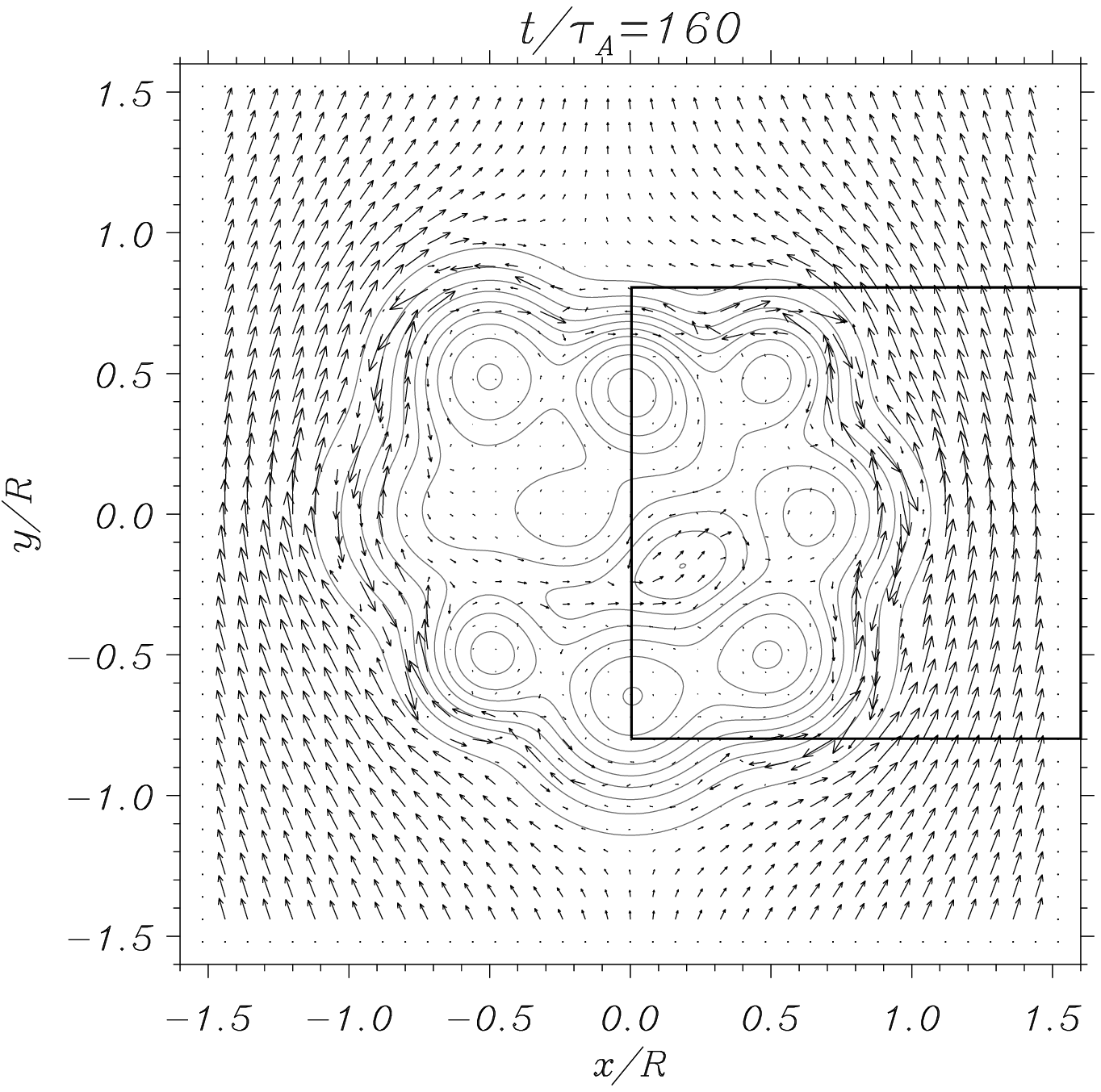}}
} \caption{ \small  Time evolution of the velocity field. The loop initially
oscillates in the $y$ direction with the global mode. Due to the mode conversion
motions localized on the magnetic surfaces develop (see the contours of
constant Alfv\'en frequency represented with continuous lines). The region marked with a square box in the bottom panel is plotted in
Figure~\ref{resonancezoom}. This figure is also
available as an mpeg animation in the electronic edition of the {\it
Astrophysical Journal}.} 
\label{resonance} \end{figure}

To explain the different behavior of the velocity at different locations in the
structure we need to understand the evolution of the whole system instead of
looking at individual points.  In Figure~\ref{contfields} the velocity
components $v_x$, $v_y$ are plotted at $t=40\,\tau_A$. Inside the loop the $v_x$
component is quite uniformly distributed while the $v_y$ component has a
complicated spatial structure which basically coincides with the spatial
distribution of the strands (see the contours). At this particular instant the
higher the density of the strand the smaller the amplitude of the $v_y$
component. On the other hand, the velocity components outside the loop have 
quite a smooth spatial distribution. The motion of the whole structure is
clearer if we represent the velocity field constructed using the $v_x$ and $v_y$
components. The result is shown in the top panel of Figure~\ref{resonance}. We
clearly see that all the strands are moving in the negative $y-$direction and
that the external medium reacts in a very organized way to the displacement of
the bundle of loops. The plasma at the bottom of the loop ($y\approx-R$) is
pushed sideways while the plasma at the top of the structure ($y\approx0.8\,R$)
tends to fill the region that has displaced downwards. This motion is the
equivalent of the kink mode in a cylindrical tube. 

At later times (see the middle and the bottom panels in Fig.~\ref{resonance})
large amplitude velocities develop specially near the external edge of the
composite loop. This is the consequence of the energy conversion between the
global mode and the Alfv\'en modes. Due to this process the global oscillation
gradually losses its energy and its amplitude is attenuated in time (see for
example the length of the arrows at the centers of the individual strands in
Fig.~\ref{resonance} at $t=160\,\tau_A$). This explains the attenuation of $v_y$
found in Figure~\ref{signalplotint}. On the other hand, the amplitude of the
Alfv\'en modes increases at the resonant layers where the energy conversion
takes place (see the large arrows at the loop boundaries in
Fig.~\ref{resonance}, bottom panel). We already found this behavior in
Figure~\ref{signalplotres}. 

\begin{figure}[!ht]
\center{
\resizebox{7cm}{!}{\includegraphics{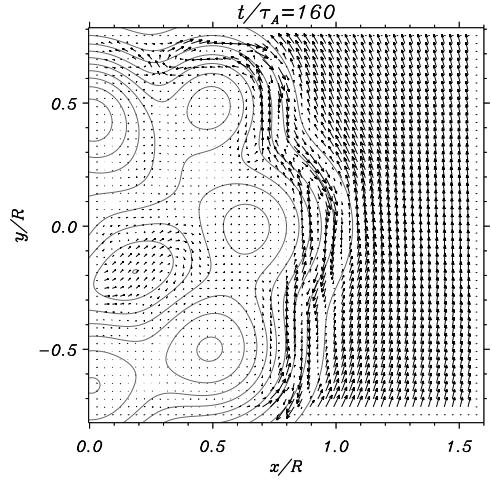}}
} \caption{ \small  Detail of the velocity field at $t=160\,\tau_A$, see square
box in the bottom panel of 
Figure~\ref{resonance}.}
\label{resonancezoom} \end{figure}

\begin{figure}[!ht]
\center{
\resizebox{7cm}{!}{\includegraphics{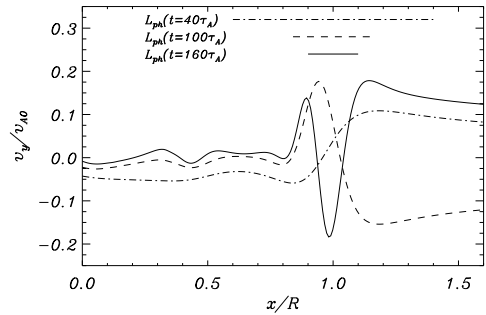}}
}

\caption{\small Plot of $v_y$  as a function of $x$ ($y=0$) at $t=40\,\tau_A$   
(dot-dashed line), $t=100\,\tau_A$ (dashed line), and $t=160\,\tau_A$
(continuous line). The lengths of the lines
plotted in the legend correspond to the phase mixing lengths, $L_{\rm ph}$,
predicted by equation~(\ref{lph}) at the position $x=R$ ($y=0$).} \label{resonancevycut} \end{figure}

Since the Alfv\'en frequency changes with position the Alfv\'enic motions get
out of phase very quickly due to phase mixing. The consequence of this process
is visible at the external boundaries where strong shear motions develop. To see
these motions in detail we have concentrated on the small region marked with a
square box in Figure~\ref{resonance} bottom panel. The result is plotted in
Figure~\ref{resonancezoom}. We see that the field is organized and that near the
edge of the loop the velocity vectors are aligned with the contours of constant
Alfv\'en frequency (see the continuous lines). It is also clear that on
neighboring magnetic surfaces motions are basically in opposite directions.
This is even clearer in Figure~\ref{resonancevycut} where we have plotted a
cut of $v_y$ at $y=0$ (see the continuous line). This figure also shows how this
component of the velocity field (which is mainly parallel to the magnetic
surfaces around $x=R$) evolves with time. As expected from the phase mixing
process, the typical spatial scales of the shear motions decrease with time. The
amplitude of the velocity decreases in the internal part of the structure while
it develops short wavelengths around the resonant positions. We find that the
typical wavelengths agree with the phase mixing lengths, $L_{\rm ph}$, predicted
by  equation~(\ref{lph}) (compare the lengths of the lines in the legend of the
plot with the wavelengths of $v_y$ at $x=R$).

At this point it is worth noticing  that due to  mode coupling Alfv\'en modes
are excited everywhere in the structure even at locations far from the
resonance. The excitation of these modes is already known, specially in driven
problems \citep[see for example][]{mann,tirry97,goossde}, but there are still
several questions about these modes that need to be addressed, as for example
the amount of energy that is deposited in the Alfv\'en modes. In any case, far
from the resonance, the amplitude of these modes is smaller that the amplitude
of the global mode (on average). We see in Figure~\ref{signalplotint} (bottom
panel) that for this internal position  the amplitude of the global oscillation,
in the $v_y$ component, is around 1.5 times the amplitude of the Alfv\'en modes
(the power of the peaks, proportional to the square of the amplitude, are around
24 and 10). A more detailed analysis, out of the scope of this work, is
required.

\subsection{Energy distribution}

\begin{figure}[!ht]
\center{
\resizebox{6.cm}{!}{\includegraphics{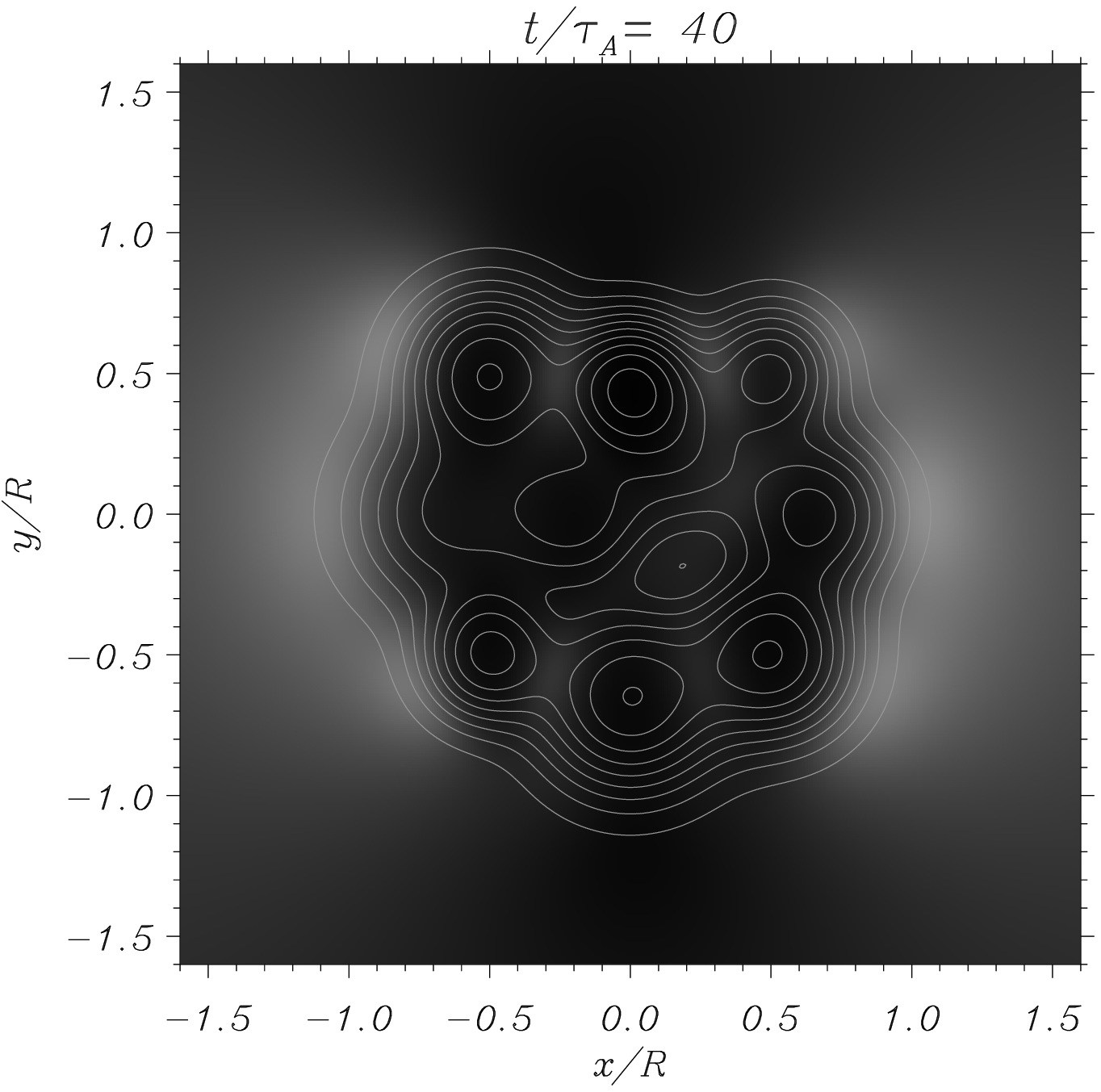}}
\resizebox{6.cm}{!}{\includegraphics{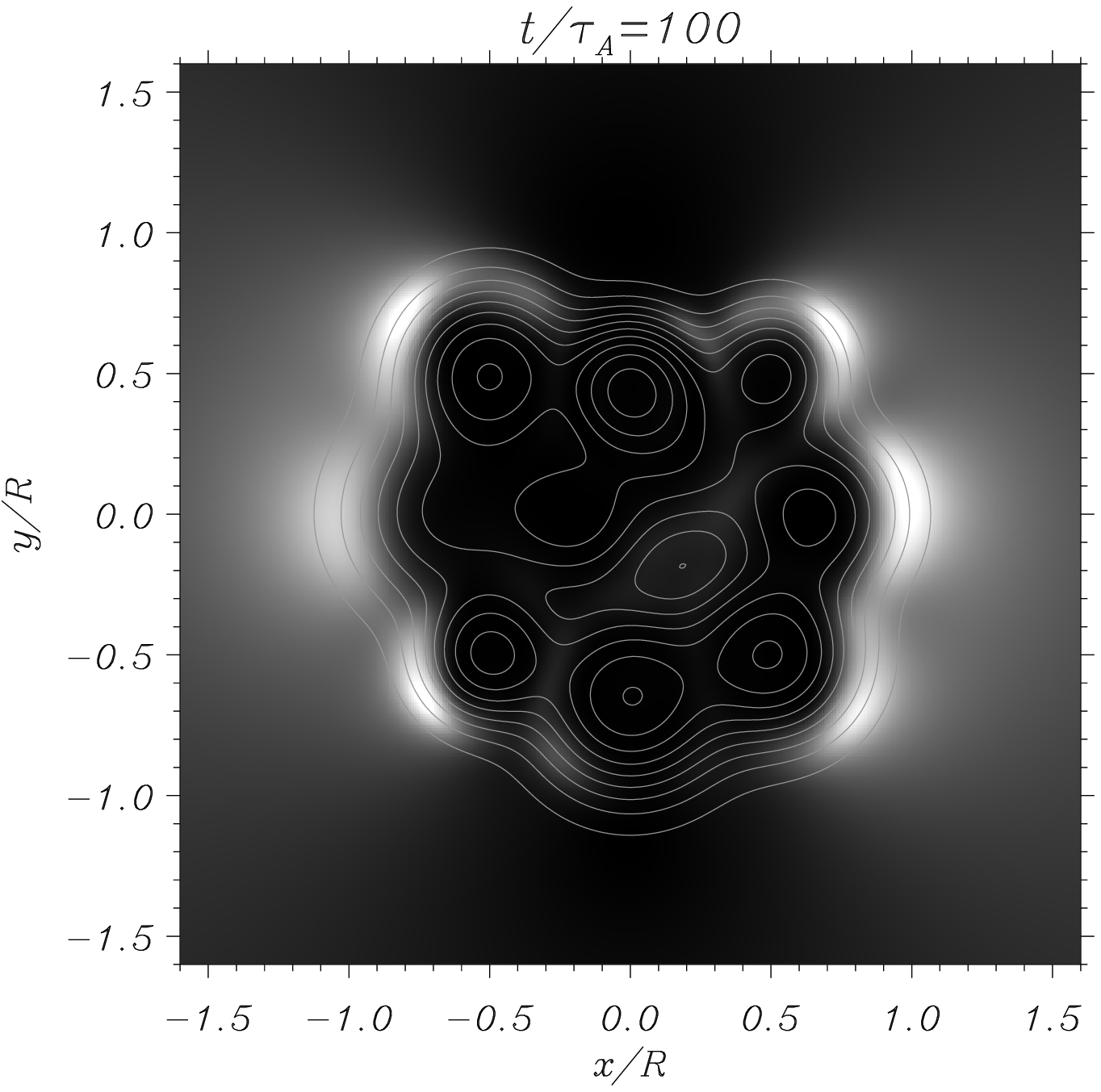}}
\resizebox{6.cm}{!}{\includegraphics{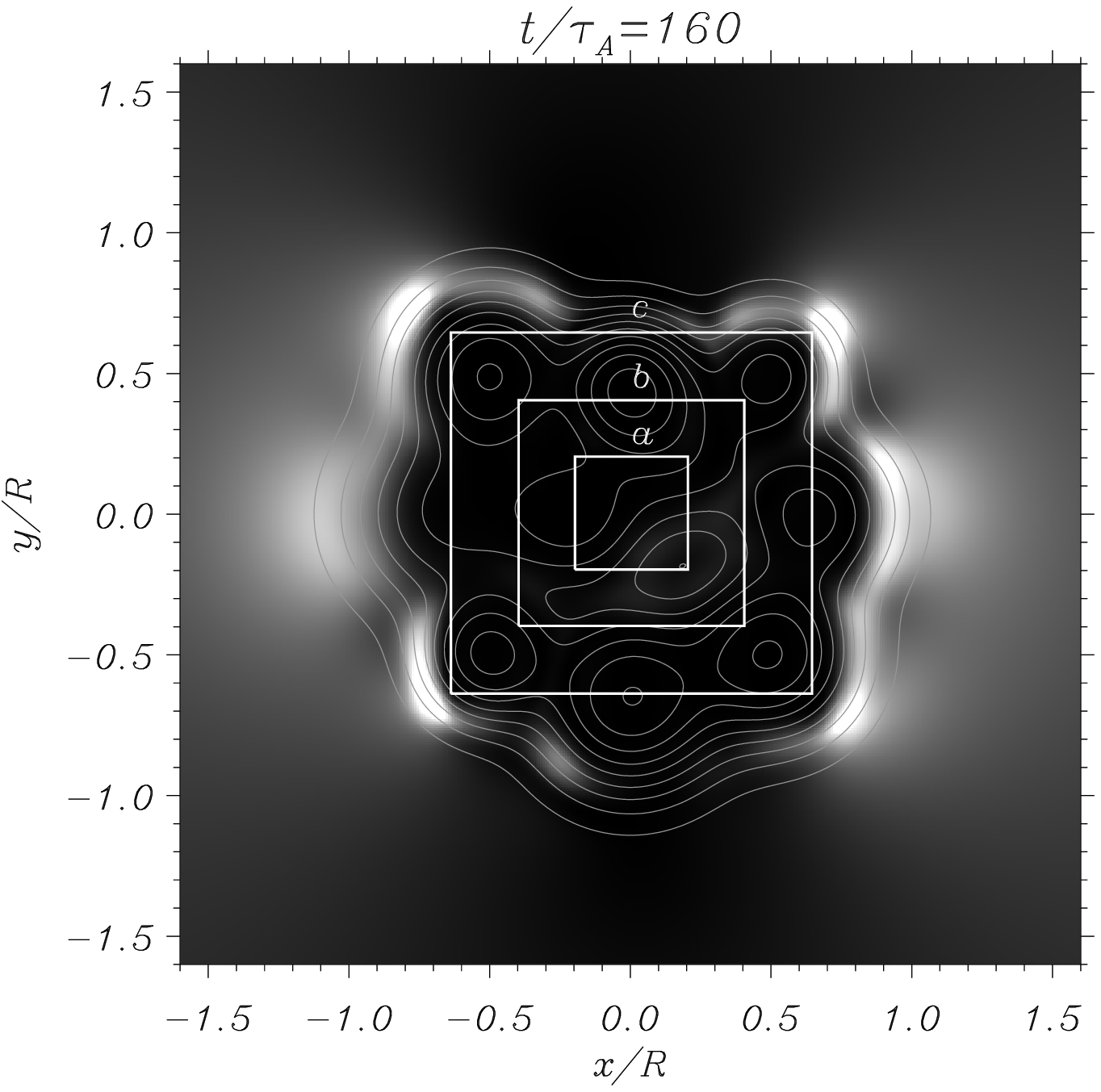}}
}
\center{\hspace{0.5cm}
\resizebox{6.cm}{!}{\includegraphics{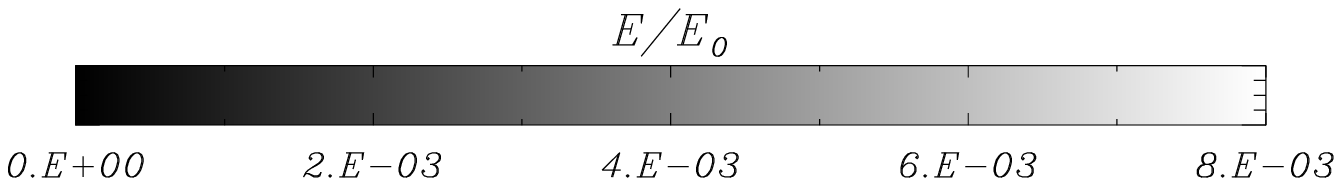}}
}
\caption{ \small  Time evolution of the energy distribution for the same
simulation as in Figure~\ref{resonance}. The three
square boxes labeled with $a$, $b$, and $c$ in the bottom panel mark the regions
where the velocity averages
have been computed to determine the damping time (see Fig.~\ref{tdaver}). This figure is also
available as an mpeg animation in the electronic edition of the {\it
Astrophysical Journal}.}
\label{resonanceenerg}  \end{figure}

\begin{figure}[!ht]
\center{
\resizebox{7cm}{!}{\includegraphics{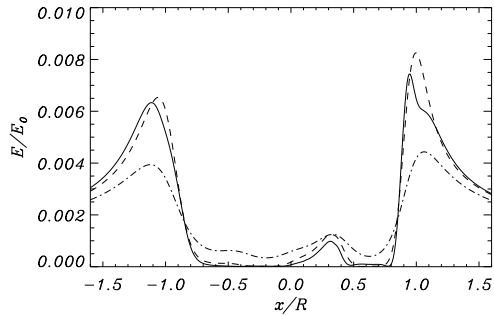}}
}
\caption{\small Wave energy density, calculated using equation~(\ref{eqener}), as
a function
of $x$ ($y=0$) for three different times (same as in Fig.~\ref{resonanceenerg},
the same notation as in Fig.~\ref{resonancevycut} has been used for the lines). 
The energy concentrates around the external edges of the loop (some part also
concentrates inside the structure, around $x=0.3R$). The
width of the energy peaks around the resonant layers decreases with time. Note
also the decrease of energy in the internal part of the loop.
} 
\label{resonanceenergcut} \end{figure}

The process of mode conversion is even clearer if we study the energetics of
the system. For this reason we now focus on the wave energy density. Since in
our model we adopt the zero$-\beta$ approximation there are only contributions
from the kinetic and magnetic energy to the total wave energy,
\begin{eqnarray}\label{eqener}
E&=&\frac{1}{2}\left[\rho_0\left(v_x^2+v_y^2\right)
+\frac{1}{\mu}\left(b_x^2+b_y^2+b_z^2\right)\right]. \end{eqnarray} This
quantity is represented in Figure~\ref{resonanceenerg} at different times (same
as in  Fig.~\ref{resonance}). The plots show that the system evolves from a
situation where the energy is more or less uniformly distributed to a state
in which it is concentrated around the resonant layers. The wave energy is
spatially distributed in such a way that it follows the irregular geometry of
the loop edge (see the bright areas on the left and right loop boundaries).  Out
of the resonances the energy decreases with time, see the interior points, except
for the area around $x=0.2R$, $y=-0.2R$ where we can find some energy for large
times. It should
be also noted  that the system evolves so that the resonance energy width
decreases with time. This behavior of the wave energy at the resonance has been
described in detail by \citet{mann} \citep[see also the equivalent results in a
cylindrical inhomogeneous loop, Fig.~8, in][]{terr06a}. To see these effects
more clearly we have concentrated on a slice at $y=0$ (see
Fig.~\ref{resonanceenergcut}) and have plotted the wave energy density at
different times. This figure shows that the energy width decreases with time at
the external boundaries, basically at $x=R$ and $x=-R$. We also see the peak
around  $x=0.3R$, and the energy decrease with time in the range $-0.8<x/R<0.1$
and $0.5<x/R<0.7$.

\subsection{Resonant magnetic surfaces}

It is quite evident from the velocity and energy analysis where the energy
conversion takes place. However, we can be more precise in the determination of
the resonant magnetic surface or surfaces. Once we know the frequency of the
global mode the location of the resonances is basically where this frequency
coincides with the local Alfv\'en frequency. In Figure~\ref{density} contours of
the Alfv\'en  frequency, $\omega_{\rm A}$, are represented together with the
global frequency, see the contours with thick lines. In this plot we find that
there are two resonant magnetic surfaces. One surface is located at the external
edge of the composite loop and the other resonant magnetic surface is located
inside the structure (see the small hole around $x=0.2R$, $y=-0.2R$). It is
precisely at these locations where the energy maps show enhanced energy, see for
example Figure~\ref{resonanceenerg} bottom panel, and also where the highest
velocity amplitudes are found, see Figure~\ref{resonance}. Thus, for this particular
multi-stranded configuration, resonant absorption not only takes place at the
external boundaries, but some energy is also deposited in the internal part of
the composite structure.

\subsection{Damping time}

\begin{figure}[hh]
\center{
\resizebox{7cm}{!}{\includegraphics{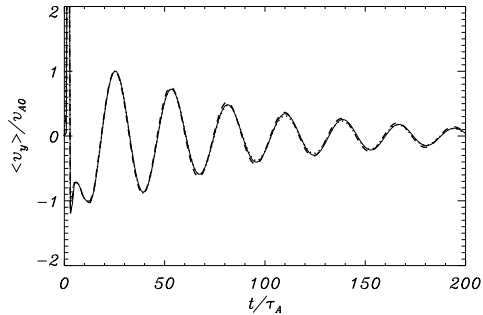}}
}
\caption{ \small  Time evolution of the spatially averaged $v_y$ component as a 
function of time. The continuous, dot-dashed, and dotted lines
correspond to the (normalized) averages in the square regions marked with labels
$a$, $b$, and $c$ in Figure~\ref{resonanceenerg} (bottom panel), respectively. The profile of
the three signals is almost the same since the loop is oscillating with the
global mode. The damping per period is $\tau_d/P=2.3$.} \label{tdaver} \end{figure}

Due to the simultaneous excitation of the quasi-mode and the local Alfv\'en
modes, the determination of the damping rates of the loop oscillation can be
quite difficult (see for example the time series in Fig.~\ref{signalplotint}).
However, there is a simple way to estimate the damping time. We have taken
uniformly distributed regions inside the loop and have added up the velocity
field at a given instant. This basically averages out the contribution of the
local Alfv\'en modes since these modes develop small spatial scales. At the same
time the average enhances the coherent  behavior of the global mode, having a
large spatial length. In Figure~\ref{tdaver} we have represented the averaged
signal as a function of time for three regions inside the loop (see the boxes
labeled with letters $a$, $b$, and $c$ in Figure~\ref{resonanceenerg} bottom
panel). We see that we get almost the same time dependence for the different
regions in the loop, indicating that the global mode is dominant everywhere
inside the loop. Although the structure is quite inhomogeneous and irregular the
damping time is basically the same everywhere. Now from the averaged signal we
can calculate the damping time using the same method as in the comparison with
the cylindrical tube, i.e. by fitting an exponentially decaying function. We
finally find that the damping per period is $\tau_d/P=2.3$.

\section{Discussion and Conclusions}\label{concl}

We have studied the mechanism of resonant absorption by solving the
time-dependent problem of the excitation of oscillations in a complicated
multi-stranded coronal loop. We have shown that the mode conversion takes place
in quite irregular geometries like the one studied in this paper and that
regular magnetic surfaces (considered in previous works) are not necessary for
this mechanism to work efficiently. This suggests that resonant absorption is 
quite a robust damping mechanism. Although we have analyzed a particular system,
the behavior found in the present equilibrium is expected to be quite generic of
inhomogeneous plasma configurations. Since inhomogeneity is certainly present in
coronal loops, the resonant coupling between fast and Alfv\'en modes can hardly
be avoided in a real situation. For this reason, resonant absorption seems to be
quite a natural damping mechanism.

The fact that the loop is not monolithic does not affect much the global
oscillatory properties. Although the loop is composed of different strands we
find a dominant frequency almost everywhere in the structure, indicating a
global motion. It is interesting to compare the frequency of this global mode
or  quasi-mode with the frequency of the equivalent homogeneous loop with radius
$R$ and the same mass. Using equation~(\ref{totdenscyl}) we find that the
equivalent cylindrical loop should have an internal density $\rho_{\rm
in}=0.67\rho_{00}$. The kink mode period for a cylindrical tube with this
internal density is $P=28.4\,\tau_{\rm A}$. This value is in good agreement with
the period of the quasi-mode estimated from the numerical simulations of the
multi-stranded model, which is around  $P=28.0\,\tau_{\rm A}$. This indicates
that the internal structure does not change much the global behavior of the loop
\citep[see also][]{arr07}. Nevertheless, it may have some effect on the location
of the energy deposition. We have found that in our model there is also energy
conversion inside the loop, although it is small compared with the energy at
resonances in the external layers.

Although at any time and at any position both the quasi-mode and the local
Alfv\'en modes are excited, we have been able to estimate the damping time of the
quasi-mode by performing averages of the dominant velocity component in different
regions of the loop. The average eliminates the local Alfv\'en modes and enhances
the global mode and is basically what the observations provide, a sort of
integration along the line of sight (but in the displacement instead of the
velocity).

On the other hand, it must be noted that a system of $N$ tubes, like the one
studied in this paper, is expected to have a large number of eigenmodes. For
example, in a configuration with  just two loops \citet{luna07} have shown that
there are four kink-like eigenmodes, and that an external disturbance in general
excites these four modes. However, these authors have shown that since their
frequencies are very similar it is very difficult to distinguish between the
different eigenmodes in a time-dependent study. For this reason, the
interpretation of the global mode found in our multi-stranded loop as a sum of
different eigenmodes with similar frequencies and similar damping times needs to
be considered. This could explain the small differences in the period and
damping time of the velocity averages in different regions of the structure. A
detailed analysis of this issue is out of the scope of this paper but it is
clear that the calculation of the eigenmodes of complicated configurations is
very important. In this regard, analytical studies based on scattering theory
will allow us to calculate the eigenmodes of a system of $N$ tubes \citep[][in
preparation]{luna08} and to make progress in this direction.

Finally, it must be noted that we have concentrated on the linear regime. If the
amplitude of the oscillations becomes large in the inhomogeneous layer due to
mode conversion, non-linear terms might be important and the efficiency of
resonant absorption can be altered. However, diffusion or viscosity may prevent
the development of large amplitudes around the resonant layer. In addition,
there are some results that indicate that in the non-linear regime the heating
at the resonant layers may produce significant changes in the equilibrium
configuration \citep[see][]{ofetal98,klim}. Hence, the initial value problem
needs to be studied using the full non-linear equations, which basically means
that, since the Fourier analysis is no longer possible in the $z-$direction, the
problem is three-dimensional. Due to the grid resolution requirements in the
layers the three-dimensional problem has a high computational cost. The
numerical study presented here for the two-dimensional problem is a preliminary
step to investigate resonant absorption in more realistic three-dimensional
models including, for example, twisted or tangled magnetic fields.

\acknowledgements  J. Terradas is grateful to the Spanish Ministry of Education
and Science for the funding provided under the Juan de la Cierva program.
Funding provided under grants AYA2006-07637, of the Spanish Ministry of
Education and Science, and PRIB-2004-10145 and PCTIB-2005GC3-03, of the
Conselleria d'Economia, Hisenda i Innovaci\'o of the Government of the Balearic
Islands, is also acknowledged.

\end{document}